\newcommand{\HII}{\mbox{H\hspace{0.2em}{\scriptsize II}}}
\newcommand{\al}{\mbox{$^{26}$\hspace{-0.2em}Al}}
\newcommand{\fe}{\mbox{$^{60}$Fe}}
\newcommand{\Msol}{\mbox{M$_{\sun}$}}

\newcommand{\yal}{\mbox{$Y_{26}^{\rm O7~V}$}}
\newcommand{\yfe}{\mbox{$Y_{60}^{\rm O7~V}$}}
\def\MeV{\mbox{Me\hspace{-0.1em}V}}

\documentclass{aa}
\usepackage{epsfig}
\begin{document}

\thesaurus{10(08.01.1;08.05.01;08.19.4;10.15.1;13.07.02)}
\title{Gamma--ray line emission from OB associations and young open 
clusters: I. Evolutionary synthesis models}

\author{M.~Cervi\~no$^{1,2}$, 
        J.~Kn\"odlseder$^{2,3}$, 
        D.~Schaerer$^1$, 
        P. von Ballmoos$^2$, and
        G. Meynet$^4$}
\institute{$^1$Observatoire Midi--Pyr\'en\'ees, 14, avenue Edouard Belin, 
               31400 Toulouse, France \\
           $^2$Centre d'Etude Spatiale des Rayonnements, CNRS/UPS, B.P.~4346,
               31028 Toulouse Cedex 4, France \\
           $^3$INTEGRAL Science Data Centre, Chemin d'Ecogia 16, 1290 Versoix,
               Switzerland \\
           $^4$Observatoire de Gen\`eve, CH--1290 Sauverny, Switzerland}

\offprints{Miguel Cervi\~no}
\mail{mcervino@obs-mip.fr}
\date{{\bf 13-Oct-2000 Accepted for A\&A main Journal}}
\authorrunning{M.~Cervi\~no et al.}
\titlerunning{$\gamma$--ray line emission from young associations}      
\maketitle

\begin{abstract}

We have developed a new diagnostic tool for the study of gamma--ray
emission lines from radioactive isotopes (such as \al\ and 
\fe) in conjunction with other multi--wavelength observables of 
Galactic clusters, associations, and alike objects.  
Our evolutionary synthesis models are based on the
code of \cite{CMH94}, which has been updated to
include recent stellar evolution tracks, new stellar atmospheres for
OB and WR stars, and nucleosynthetic yields from massive stars during
hydrostatic burning phases and explosive SN~II and SN~Ib events.

The temporal evolution of \al\ and \fe\ production, the equivalent yield 
of \al\ per ionising O7~V star (\yal), and other observables are 
predicted for a coeval population. 
The main results are:
\begin{itemize}
\item The emission of the \al\ 1.809 MeV line is characterised by 
      four phases:
      stellar wind dominated phase ($\la$ 3 Myr),
      SN~Ib dominated phase ($\sim$ 3--7 Myr),
      SN~II dominated phase ($\sim$ 7--37 Myr),
      and exponential decay phase ($\ga$ 37 Myr).
\item The equivalent yield \yal\ is an extremely sensitive age 
      indicator for the stellar population which can be used to 
      discriminate between Wolf--Rayet star and SN~II \al\ nucleosynthesis 
      in the association.
\item The ratio of the \fe/\al\ emissivity
      is also an age 
      indicator that constrains the contribution of explosive 
      nucleosynthesis to the total \al\ production.
\end{itemize}

We also employed our model to estimate the steady state 
nucleosynthesis of a population of solar metallicity.
In agreement with other works, we predict the following relative 
contributions to the \al\ production:
$\sim 9 \%$ from stars before the WR phase, 
$\sim 33 \%$ from WR stars, 
$\sim 14 \%$ from SN~Ib, 
and $\sim 44 \%$ from SN~II.
For \fe\ we estimate that $\sim 39 \%$ are produced by SN~Ib while 
$\sim 61 \%$ come from SN~II.
Normalising on the total ionising flux of the Galaxy, we predict total 
production rates of 1.5 \Msol\ Myr$^{-1}$ and 0.8 \Msol\ 
Myr$^{-1}$ for \al\ and \fe, respectively.
This corresponds to 1.5 \Msol\ of \al\ and 1.7 \Msol\ of \fe\ in
the present interstellar medium.

To allow for a fully quantitative analysis of existing and future
multi--wavelength observations, we propose a Bayesian approach that 
allows the inclusion of IMF richness effects and observational 
uncertainties in the analysis.
In particular, a Monte Carlo technique is adopted to estimate 
probability distributions for all observables of interest.
We outline the procedure of exploiting these distributions by 
applying our model to a fictive massive star association.
Applications to existing observations of the Cygnus and Vela regions
will be discussed in companion papers.

\end{abstract}

\section{Introduction}

OB associations and young open clusters are the most active
nucleosynthetic sites in our Galaxy.  The combined activity of stellar
winds and core--collapse supernovae ejects significant amounts of
freshly synthesised nuclei into the interstellar medium.  Among those,
radioactive isotopes, such as \al\ ($\tau=1.04 \times 10^6$~yr) or
\fe\ ($\tau=2.07 \times 10^6$~yr), may eventually be observed by
gamma--ray instruments through their characteristic decay--line signatures.
Their observation presents direct evidence of recent nucleosynthesis 
activity, which can be used as a powerful diagnostics tool for studies 
of present galactic activity.

Galactic 1.809 \MeV\ gamma--ray line emission attributed to the radioactive
decay of \al\ has been observed by numerous gamma--ray telescopes, and the
detailed mapping of the emission distribution by the COMPTEL telescope has
clearly identified massive stars as the source of this radio--isotope (see
\nocite{prantzos96}Prantzos \& Diehl 1996 for a review).  The most
convincing evidence for a massive star origin comes from the close
resemblance between the 1.809 \MeV\ and galactic free--free emission, which
links \al\ nucleosynthesis to the O star population (Kn\"odlseder et
al.~1999a\nocite{knoedl99a}; Kn\"odlseder 1999\nocite{knoedl99}).  Since in
general a variety of distinct massive star populations of different ages,
sizes, and metallicities contribute to the observed intensities along a
line of sight, this indicates that the average properties of the
populations are related.  However, the correlation between 1.809 \MeV\ and
free--free emission also holds for regions far from the Galactic centre
where only few massive star associations contribute to the observed
emissions.  Examples are the Cygnus and the Vela regions where localised
1.809
\MeV\ emission enhancements coincide spatially with maxima of 
free--free radiation, showing the same relative intensities as the 
Galaxy as a whole
(Kn\"odlseder et al.~1999a\nocite{knoedl99a}).

To fully exploit in a quantitative manner such existing data and in
preparation of the upcoming {\em INTEGRAL} gamma--ray satellite mission
the development of new interpretation tools is necessary.  We here
present the first results from our modified time--dependent
multi--wavelength evolutionary synthesis models.  A similar model was
recently presented by \cite{Pl00}.  To model the
gamma--ray luminosities the nucleosynthetic production of the
long--lived radio--isotopes \al\ and \fe, has been included in our
multi--wavelength code (Cervi\~no \& Mas-Hesse 1994\nocite{CMH94}, Cervi\~no et al.\
2000)\nocite{CMHK00}.  These isotopes give rise to the 1.809 \MeV\ (for \al) and
1.173 \MeV\ and 1.333 \MeV\ (for \fe) gamma--ray lines respectively.
The model properly accounts for the accumulation of radioactive
elements and their respective decay times.  Results from
state--of--the--art stellar atmospheres of massive stars are included to
accurately predict the ionising fluxes from these stars, which are at
the origin of thermal free--free emission. Together with the other
synthesised observables this provides a versatile tool for gamma--ray to
radio analysis of massive star forming regions.

For comparisons with individual Galactic star forming regions (e.g.\
OB associations, clusters, H~{\sc ii} regions) or ensembles of such
objects it is not only imperative to model the temporal evolution of
their properties. The effects of small number statistics of the
massive star population must also be taken into account (Cervi\~no et
al.\ 2000b\nocite{CLC00}).  Various studies, including the present one, treat such
effects by means of Monte Carlo simulations (e.g.\ Cervi\~no \&
Mas-Hesse 1994\nocite{CMH94}, McKee \& Williams 1997\nocite{MKW97}, Oey \& Clarke 1998\nocite{Oetal98}).  Finally
for a fully quantitative and objective confrontation with observables
an additional step is performed here, to our knowledge for
the first time in this context.  The observational constraints (e.g.\
a known number of stars of given spectral types, derived age, distance
etc.) and their uncertainties are included in a Bayesian approach 
providing probability distributions for all derived properties.

Section 2 describes the ingredients of our synthesis code.  The main
predictions from our models are presented in Section 3.  Uncertainties
are briefly discussed in Sect.\ 4.  Our Bayesian approach to model
realistic stellar populations is presented in Sect.\ 5.  The main
conclusions are summarised in Sect.\ 6.

\section{Evolutionary synthesis model}

\subsection{Method}
\label{sec:method}

The starting point for our modelling effort is the evolutionary synthesis
code of \cite{CMHK00}, which predicts the time--dependent multi--wavelength
energy distribution of a population of discrete stars from radio
wavelengths up to the X--ray domain.  For this work, we have updated the
atmosphere models for massive stars using the CoStar models (Schaerer \& de
Koter 1997\nocite{SK97}), and we included atmosphere models for the
Wolf--Rayet (WR) phase from \cite{Schetal92} following the prescriptions of
Schaerer \& Vacca (1998)\nocite{SV98}.  In order to predict gamma--ray
luminosities, we have included chemical yields for the radioactive isotopes
\al\ and \fe\ that may either be produced during hydrostatic
nucleosynthesis in the interiors of massive stars, or during explosive
nucleosynthesis in supernova explosions (cf.~Section \ref{sec:yields}).

The calculations have been done for two different star formation laws
in order to explore the extreme cases of an instantaneous burst (IB)
and of a constant star formation rate (CSFR).  For the IB model, an
initial population of coevally formed stars has been created based on
a Monte Carlo method.  Using a power--law initial mass function (IMF)
of slope $\Gamma$ as probability density function\footnote{The
Salpeter IMF has a slope of $\Gamma=-1.35$ in this prescription.}, we
randomly created initial stellar masses within the interval $2-120$
\Msol\ until the total number of stars reaches a predefined limit.
The evolution of each star is then calculated using the Geneva
evolutionary tracks (see below) in time steps of $10^5$ years up to an
age of typically $50$ Myr.  At each time step the spectral energy
distribution and the ejected \al\ and \fe\ yields are computed for
each individual star.  The evolution of spectral types is also
followed in order to predict the number of O and WR stars in the
population.  Stars that end their lives during a time step are counted
as supernova explosions (as far as they are more massive than $8$
\Msol), and are removed from the population for the next time step.
Summing the contributions from all individual star results then in
predictions for the entire population.

\subsection{Evolutionary tracks}

The evolution of the stars in the population is followed using the
non--rotating stellar tracks from \cite{MAPP97} (hereafter MAPP97)
for stars with initial mass $M_{\rm ini} \ge 25$ \Msol,
\cite{Meyetal94} for $15 \le M_{\rm ini}/\mathrm{M}_\odot \le 20$, and
\cite{Scha92} otherwise.  Solar metallicity tracks are used for this
work since we are interested in predictions for star clusters and OB
associations in the solar  neighbourhood, but in future we plan to
extend the calculations also to other metallicities.  The possible
alterations when rotation is taken into account in the stellar models
are briefly discussed in Sect.\ \ref{sec:uncertain}.  Rotating stellar
models will be included when complete tracks covering all relevant
evolutionary phases will become available.

The stellar tracks we used have been calculated for enhanced mass loss
during the massive star evolution until the end of the WNL phase.  This
prescription leads to an improved agreement between predictions and several
observed WR properties; in particular, these models can account for the
variation of the number ratio of WR to O type stars as a function of the
metallicity in zones of constant star formation rate (Maeder \& Meynet
1994\nocite{MM94}).  The relative populations of WN and WC stars observed
in young starburst regions are also better reproduced when models with high
mass loss rates are used (Meynet 1995\nocite{GM95}; Schaerer et al
1999\nocite{Schae99}).

The models we are using predict a lower initial mass limit of
25 \Msol\ for the formation of WR stars. Uncertainties due to this
mass limit will be discussed in Section \ref{sec:uncertain}.
In order to avoid numerical inconsistencies and an unrealistic behaviour
of interpolated tracks around the WR mass limit, we have constructed an
artificial track at 25.01 \Msol\ going through the WR phases. 
This track together with the published 25 \Msol\ track, which does not enter 
the WR phase, allows smooth interpolations in this mass range.

\subsection{Chemical yields}
\label{sec:yields}

\begin{figure*}
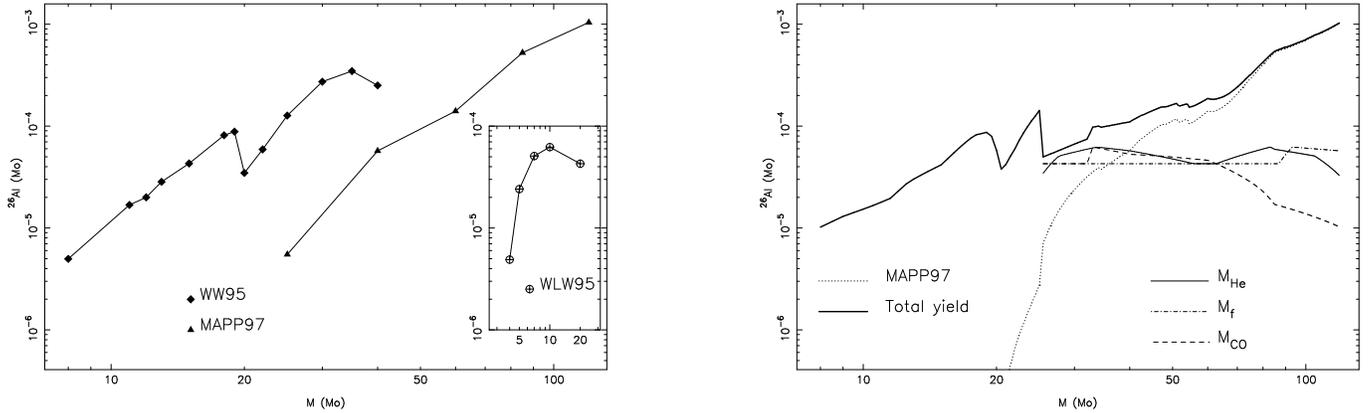

\epsfig{file=10196a.f1, angle=270,  width=8cm}
\hfill
\epsfig{file=10196b.f1, angle=270,  width=8cm}
\caption{\al\ yields used in this work.
    {\bf Left panel:}
         \al\ yields as function of initial stellar mass for
         Wolf--Rayet stars (Meynet et al.~1997, MAPP97, triangles) and
         type II supernovae (Woosley \& Weaver 1995, WW95: diamonds).
         The inset shows \al\ yields for Helium stars as function of
         the initial He mass (Woosley et al.\ 1995, WLW95).
   {\bf Right panel:}
         \al\ yields as function of initial stellar mass after
         combining the nucleosynthesis models with the evolutionary
         tracks (see text).  For type Ib/c supernovae, 3 link
         parameters ($M_{\rm He}$: thin solid line, $M_{f}$:
         dashed--dotted, $M_{\rm CO}$: dashed) have been explored,
         which lead to comparable \al\ yields.  The resulting total
         yield, obtained using $M_{\rm CO}$, is shown by the thick
         solid line }
\label{fig:alyields}
\end{figure*}

The prediction of chemical yields for massive stars depends critically
on the assumed stellar physics, such as the treatment of convection,
rotation, mass--loss, and the final supernova explosion.  Being aware
of these uncertainties, we do not intend to predict the yields of \al\
and \fe\ to better than within a factor of 2 or so (e.g.~Prantzos \&
Diehl 1996\nocite{prantzos96}; Woosley \& Heger 1999\nocite{woosley99}), 
but we merely want to identify the main
characteristics of nucleosynthesis in a massive star population.
However, by comparing the models to real massive star populations, one
can inverse the problem and try to learn something about massive star
nucleosynthesis, and possibly better constrain the theoretical models.
In this sense, the employed nucleosynthesis yields can be seen as a
hypothesis, which can be verified by comparison to observations.

Two different sites of nucleosynthesis must be taken into account for
the prediction of \al\ and \fe\ yields from a massive star
association.  First, the H--burning in the core of the stars may
produce appreciable amounts of \al\ that may appear at the stellar
surface as an effect of both internal mixing and removal of the
external layers by stellar winds. This \al\ can then be ejected into
the interstellar medium by the stellar winds.  Second, when the star
explodes in a supernova event, both \al\ produced during the post
H--burning phases and that synthesised at the time of the explosion
are then expulsed into the interstellar medium.

To obtain the yields from the population synthesis code, one needs a
series of different initial mass stellar models computed with the same
physical ingredients. Since the Geneva tracks stop before the
presupernova stage and give no predictions concerning the explosive
nucleosynthesis, we must complement these data with the yields of \al\
and \fe\ ejected during the supernova event.

\subsubsection{Stellar winds}

\begin{figure*}
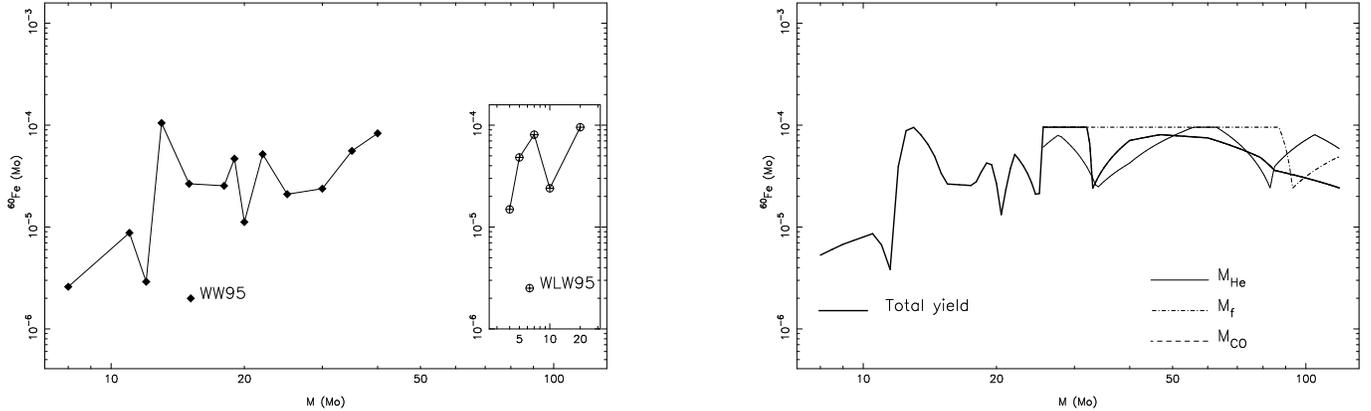

\epsfig{file=10196a.f2, angle=270,  width=8cm}
\hfill
\epsfig{file=10196b.f2, angle=270,  width=8cm}
\caption{\fe\ yields used in this work.
    {\bf Left panel:}
         \fe\ yields as function of initial stellar 
         mass for type II supernovae (Woosley \& Weaver 1995, WW95).
         The inset shows \fe\ yields for Helium stars as function 
         of the initial He mass (Woosley, Langer \& Weaver 
         1995, WLW95).
   {\bf Right panel:}
         \fe\ yields as function of initial 
         stellar mass after combining the nucleosynthesis models with the 
         evolutionary tracks (see text). Same symbols as in 
         Fig.~\protect\ref{fig:alyields}. 
         For $M \ge 25$ \Msol, the final yield corresponds to the yield 
         using $M_{\rm CO}$ as a linking parameter}
\label{fig:feyields}
\end{figure*}

Recent calculations of hydrostatic \al\ nucleosynthesis in mass losing stars 
have been undertaken 
by \cite{langer95} and MAPP97\nocite{MAPP97} for non--rotating single massive
stars.  Despite the different assumptions that both groups made about
mass loss and convection, both calculations lead to comparable results
(MAPP97\nocite{MAPP97}).  We here use the MAPP97\nocite{MAPP97}
calculations, which have the advantage of being consistent with the adopted 
evolutionary tracks which have extensively been compared to observations
(see Sect.~2.2).
The dependence of the total mass of \al\ ejected by WR stellar winds as a 
function of the initial mass is shown in the left panel of
Fig.~\ref{fig:alyields}.
We have assumed no \al\ ejection by stellar winds for the 20~\Msol\ stellar model.  Log--linear interpolation is used to
determine the yields at intermediate masses.

Arnould et al.~(1997)\nocite{Arnetal97} suggest that some amount of \fe\ could be ejected
by WR stellar winds. Typically, for a 60 \Msol\ star they find $\sim 10^{-10}$
\Msol\ of \fe\ ejected during the WR phase, which is well below the quantity
of \fe\ expelled in the final SN explosion (cf.\ Fig.\ \ref{fig:feyields}).
The stellar wind contribution of \fe\ can therefore safely be neglected
for our purposes.

\subsubsection{Type II supernova nucleosynthesis}

At solar metallicity, stars in the mass range $\sim 8-25$ \Msol\ will end 
their lives by core collapse, giving rise to type II supernova 
(SN~II) explosions.
Explosive \al\ and \fe\ nucleosynthesis during these events has been 
calculated by Woosley \& Weaver (1995; hereafter WW95)\nocite{WW95},
\cite{Thietal96}, and \cite{Limetal00}.
Their models differ in the treatment of the pre--supernova evolution, 
the prescription of convection, the employed nuclear reaction 
networks, and the assumed explosion mechanism.
For example, while WW95\nocite{WW95} included hydrostatic \al\ production in the 
pre--supernova phase in their models, \cite{Thietal96} only predict 
explosive nucleosynthesis yields in their models.
Additionally, WW95\nocite{WW95} added neutrino driven spallation (the so called 
$\nu$--process) in their reaction network while the others ignore this 
channel (the $\nu$--process may enhance \al\ production due to 
additional release of protons).

Of all these models, WW95\nocite{WW95} predict the highest \al\ yields; for example
\cite{Thietal96} obtain yields that are almost a factor of 10 lower. 
We here adopt the WW95\nocite{WW95} yields for \al\ and \fe, whose dependence
on the initial stellar mass are shown in the left panels of
Figs.~\ref{fig:alyields} and \ref{fig:feyields} respectively.
Note that the WW95\nocite{WW95} yields do not reach down to $8$ \Msol, the assumed
initial mass limit for SN~II.  We assume no \al\ production for 7~\Msol\ 
and perform a log--linear interpolation in the mass interval between 8 and 
11 \Msol.
Note that in the case of \fe\ the choice of the inferior mass
limit for stars undergoing core--collapse has an important impact on 
the results since the stars in this mass range may have a considerable 
contribution.

In order to assign a supernova model (calculated from stellar models
neglecting mass loss and with different limits of the convective cores) to
the adopted stellar models, we use $M_{\rm CO}$, the mass of the
Carbon--Oxygen core at the end of C--burning, as linking parameter, as
suggested by \cite{Mae92}.  This procedure is based on the hypothesis that
the relation between $M_{\rm CO}$ and the explosive nucleosynthetic yields
does not much depend on the particular set of stellar models.  In
particular for our case this should be a reasonable assumption since at the
time of the supernova explosion the main regions of \al\ and \fe\
production are inside the CO core (cf.\ WW95\nocite{WW95}).  $M_{\rm CO}$
from the evolutionary tracks was estimated from the fraction of the
convective core before the end of He burning\footnote{Corresponding to
point 42 in the tables of the tracks, where the central Helium mass
fraction is 0.1. This is justified since the subsequent evolution during
C--burning should not alter $M_{\rm CO}$}.  Although, for the tracks in
common with \nocite{Mae92}Maeder (1992), the derived values
are somewhat lower than the ones tabulated by \nocite{Mae92}Maeder (1992), the yields are only
slightly modified. For the WW95\nocite{WW95} models we use the $M_{\rm CO}$
values from
\cite{Poretal98} calculated by subtracting the amount of hydrogen and
helium in the WW95\nocite{WW95} tables from the initial mass.  The resulting \al\ and
\fe\ yields are shown in the right panels of Figs.~\ref{fig:alyields} and
\ref{fig:feyields},

\subsubsection{Type Ib/c supernova nucleosynthesis}
\label{sec:typeIb}

For stars that go through the  Wolf--Rayet phase, the above approach of 
estimating the explosive nucleosynthesis yields is not longer valid.
The mass--loss will considerably modify the structure of the star 
prior to explosion, leading eventually to a type Ib or type Ic 
supernova explosion at the end of its lifetime.
After the evaporation of the hydrogen envelope, such an object may 
closely resemble a Helium star.

\begin{figure}
\epsfig{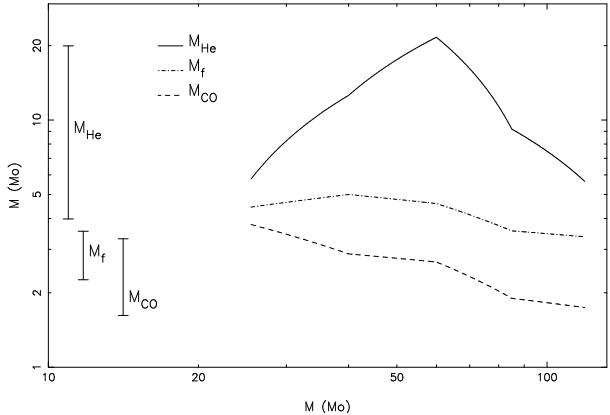}
\caption[]{$M_{\rm He}$, M$_{\rm CO}$, and $M_{\rm f}$ mass versus 
           initial stellar mass as determined from the evolutionary 
           tracks.
           The mass ranges in these three parameters covered by the 
           nucleosynthesis models of WLW95 are shown as vertical lines 
           in the left.}
\label{fig:MWLW}
\end{figure}

The only computation of explosive nucleosynthesis yields for \al\ and 
\fe\ in such events comes from Woosley, Langer \& Weaver 
(1995\nocite{WLW95}; hereafter WLW95) who calculated the explosion of 
mass losing Helium stars of initial He masses between $4-20$ \Msol.
Again, the models of WLW95\nocite{WLW95} have to be connected to the evolutionary 
tracks in our evolutionary synthesis model.
In principle there are three different ways how such a link could be 
done.
First, we could use the final mass $M_{f}$ of the MAPP97\nocite{MAPP97} 
evolutionary tracks and link them to the final masses of the WLW95\nocite{WLW95} 
models.
Second, we could use the mass of the He core $M_{\rm He}$ at the 
beginning of core He burning and connect them with the initial masses 
of the WLW95\nocite{WLW95} models.
And third, as for SN of type II,  $M_{\rm CO}$ could be used.

The three possible link parameters $M_{f}$, $M_{\rm He}$ and $M_{\rm CO}$
derived from the evolutionary tracks are shown as function of the initial
stellar mass in Fig.~\ref{fig:MWLW}.
Since $M_{\rm He}$ and $M_{\rm CO}$ are not all directly available from the 
tracks they were estimated
from the mass fraction of the convective core close to the beginning and
end of He--burning respectively\footnote{Points 23 and 42 of the tracks.}.
The $M_{\rm CO}$ values estimated in this manner are found to be $\sim$
20 -- 40 \% lower than $M_{\rm CO}$ from the stellar structure models
(Foellmi 1997\nocite{Foe97}).
As shown by Fig.~\ref{fig:MWLW} not all mass ranges overlap with
available SN Ib models of WLW95\nocite{WLW95}. Some link parameters are therefore
of limited practical use.

The \al\ and \fe\ yields resulting from the use of the different
link parameters (excluding extrapolations outside the range covered
by the WLW95\nocite{WLW95} models) is shown in the right panels of Figs.~\ref{fig:alyields} 
and \ref{fig:feyields} (values at $\ge$ 25 \Msol). 
Given the physical and numerical uncertainties in the link between 
the hydrostatic stellar models and the SN Ib calculations the variations
are considered to be small.
For masses $\ga$ 70 \Msol, the link using M$_{\rm CO}$ provides a somewhat 
lower \al\ yield than using $M_{\rm He}$. However, since these stars are
relatively rare and since stellar wind ejection exceeds the SN Ib production 
by up to an order of magnitude, the precise explosive yield in this domain is 
not crucial.
For consistency with the type II supernova yields, $M_{\rm CO}$ will thus
be used for all tracks as the link parameter in the remainder of this work.

Note also that \cite{knoedl99}, in his estimation of the global 
galactic \al\ production rate, used a mass independent, constant SN~Ib/c 
yield of $6 \times 10^{-5}$ \Msol.
This prescription is in good agreement with our more refined 
treatment, which predicts yields between $(4-6) \times 10^{-5}$ \Msol.

\section{Model predictions}
\label{sec:predictions}

We will now present the temporal evolution of some of the key 
predictions of our model, such as the supernova rate, the ionising 
flux, and the \al\ and \fe\ nucleosynthesis yields.
As indicated earlier (Sect.\ \ref{sec:method}) we consider two different 
star formation histories: a coeval population (instantaneous burst: IB)
and a constant star formation rate (CSFR).
In the present section an analytic description of the IMF is used.
Realistic populations of OB associations and young open clusters, 
with a limited number of member stars, will be discussed in Section
\ref{sec:realistic}.
We adopt a Salpeter IMF ($\Gamma=-1.35$) over the interval 
$2-120$~\Msol, 
variations of the IMF slope will be discussed in 
Sect.~\ref{sec:IMFdep}.
Recall that in both the IB and CSFR cases our normalisation yields
{\em absolute} quantities given per mass of stars formed (IB case), and
star formation rate, \Msol/yr (CSFR case).
All other predictions shown here, 
referring to relative quantities, are not affected by the adopted
normalisation.

\subsection{Supernova rates}
\label{sec:snr}

\begin{figure}[ht]
\epsfig{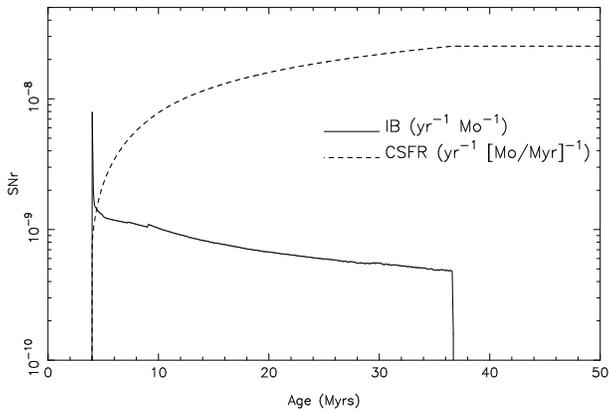}
\caption[]{Predicted temporal evolution of the supernova rate for 
  the IB (solid) and CSFR (dashed) star formation laws.}
\label{fig:SNr}
\end{figure}

The predicted supernova rates from our models are shown
in Fig.~\ref{fig:SNr}.
For the IB law, the supernova activity starts with a sharp peak 
around $\sim4$ Myr which then soon turns over into a smoothly declining 
activity, situated around 1 SN per Gyr and \Msol.
The peak is due to the fact that stars within the mass interval 
$60-120$ \Msol\ all have about the same lifetime 
(3.97, 3.95, and 4.05 Myr for a 120, 85, and 60 \Msol\ star, 
respectively), hence the stars within this mass range explode at almost 
the same moment.
The supernova activity ends around $\sim37$ Myr, when all stars more 
massive than $8$ \Msol\ have vanished. 

\begin{figure}[ht]
\epsfig{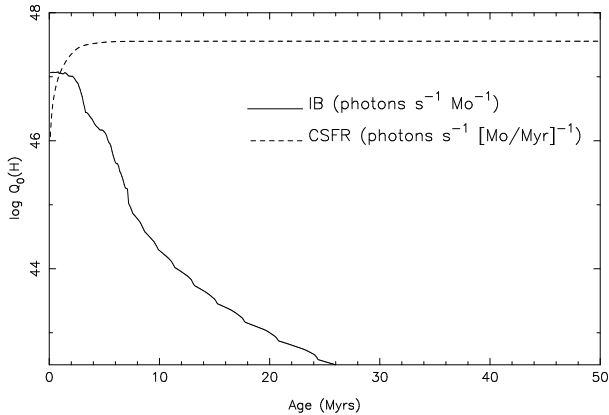}
\caption[]{Predicted temporal evolution of the ionising flux for 
  the IB (solid) and CSFR (dashed) star formation laws.}
\label{fig:Q}
\end{figure}

For the CSFR, the onset of the supernova activity is much more smooth, 
turning quickly into an almost constant rate of $\sim25$ SN per Gyr and 
\Msol/Myr\footnote{The choice of \Msol/Myr instead of the commonly 
used \Msol/yr is for illustrative purposes only.}.

\subsection{Ionising flux}
\label{sec:Q}

The evolution of the ionising flux $Q_0$, defined as the number of
photons emitted per second with wavelengths shorter than 912 \AA, is
shown in Fig.~\ref{fig:Q}.  After the onset of star formation $Q_0$
decreases with age.  In the case of an IB, the decline is very rapid,
reducing during the first 12 Myr the ionising flux by a factor of
$10^3$.  This comes from the fact that the bulk of ionising flux is
provided by stars more massive than $\sim20$ \Msol, which disappear
within only a few million years after their formation.  In the case of
a CSFR new massive stars constantly replenish the loss of ionising
photons from stars which disappear.  Since the ionising flux increases
strongly with mass, $Q_0$ reaches equilibrium more rapidly than the SN
rate.

\subsection{\al\ and \fe\ ejection rates}

\begin{figure*}[ht]
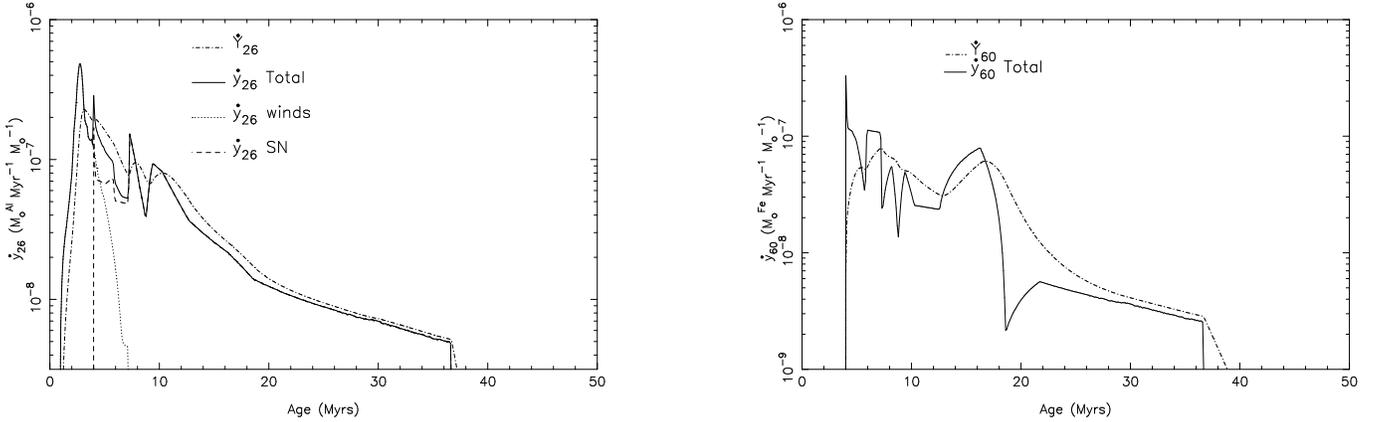

\epsfig{file=10196a.f6, angle=270,  width=8cm}
\hfill
\epsfig{file=10196b.f6, angle=270,  width=8cm}
\caption[]{Temporal evolution of the predicted \al\ (left panel) and 
           \fe\ (right panel) ejection rates for instantaneous burst
           models.  The immediate total rate is shown by the solid line.
           Contributions from hydrostatic nucleosynthesis (dotted line) and
           from explosive nucleosynthesis (dashed line) are shown.  The
           dashed--dotted line shows the emissivity as defined by Eq. (1)
           in the text.}
\label{fig:rates}
\end{figure*}

The \al\ and \fe\ ejection rates, 
$\dot{y}_{26}$ and $\dot{y}_{60}$ respectively,
defined as the mass of radio--isotopes 
ejected per Myr, and normalised to the total mass converted into stars,
are shown in Fig.~\ref{fig:rates} for IB models.
Several ejection peaks are seen in the evolution of the \al\ rates.
The first peak between $2-3$ Myr, which presents the maximum ejection 
rate during the evolution, comes from the onset of strong winds for the 
most massive stars in the population.
The next peak at $\sim4$ Myr is due to the almost simultaneous 
explosion of all stars in the mass interval $60-120$ \Msol\ as type 
Ib/c supernovae (see Section \ref{sec:snr} and Fig.~\ref{fig:SNr}).
Around $\sim5$ Myr type Ib/c supernovae start to dominate \al\ 
production, until at $7$ Myr the first type II supernovae begin to 
explode.
The enhanced \al\ production of type II SN with respect to type Ib/c 
leads then to the next peak between $7-8$ Myr.
From this age on, the \al\ ejection rate is dominated by type II 
events.
The time--dependence of the ejection rate then reflects the mass dependence 
of type II supernova yields (cf.~Fig.~\ref{fig:alyields}) 
and the slow decline of the supernova rate (cf.~Fig.~\ref{fig:SNr}).

For the \fe\ ejection rate, a complex structure is found, where the 
first peak again reflects the burst of supernova explosions around 
$\sim4$ Myr, and the remaining peaks directly reflect the dependency 
of the \fe\ yield on initial stellar mass (cf.~Fig.~\ref{fig:feyields}).
Hence the temporal structure of the \fe\ rate depends mainly on the 
details of the explosive nucleosynthesis models, and following the 
discussion about the uncertainties of these models 
(Sects.\ \ref{sec:yields}, \ref{sec:uncertain}),
the structure should not be regarded as a physical 
prediction of our model.
In particular, the broad bump between $13-18$ Myr mainly reflects 
the single high--yield point in the WW95\nocite{WW95} models at $M=13$ \Msol, and 
modifications in the yields due to changes in the assumptions about the 
stellar physics in the nucleosynthesis models (e.g.~Woosley \& Heger 
1999\nocite{woosley99}) can easily shift or remove this bump.

Emissivities (i.e. decay rates) $\dot{Y}(t)$, in units of \Msol\ per Myr,
and normalised to the total mass converted into stars, have been obtained
by integrating the ejection rates $\dot{y}(t)$ over the past history
including the radioactive decay, using
\begin{equation}
 \dot{Y}(t) = \tau^{-1} \int_{0}^t \dot{y}(t') \exp 
              \left( -(t-t') / \tau \right) dt' ,
\label{eq:cumulative}
\end{equation}
where $\tau$ is the mean life of the radioactive isotope ($\tau=1.04$ Myr
for \al\ and $\tau=2.07$ Myr for \fe).  As seen in Fig.\ \ref{fig:rates}
the emissivities are much smoother than the ejection rates, which is an
obvious result of the convolution operation given by
Eq.~\ref{eq:cumulative}.  For \al, the emissivity rises sharply between
$1-3$ Myr, which is attributed to stellar wind ejecta from massive stars.
From $\sim3$ Myr on, the emissivity decreases continuously with a secondary
maximum around $7-8$ Myr due to the onset of type II supernovae mass
ejections.  The trend of decreasing \al\ rate with increasing age should be
a generic feature for \al\ nucleosynthesis in massive star associations, at
least for a Salpeter IMF, independently of the uncertainties in the
nucleosynthesis models.  The reason is that hydrostatic \al\ production,
which dominates \al\ nucleosynthesis in WR stars and SN~II, decreases
generally with decreasing initial mass since the number of seed nuclei
becomes smaller.  Also, the SN rate, which defines the number of ejection
events within a time interval, decreases with time
(cf.~Fig.~\ref{fig:SNr}).  However, the level of the first maximum
(i.e.~the WR--peak) with respect to the second maximum (i.e.~the SN--peak)
may depend on details of the nucleosynthesis calculations.  Also, the age
at which the second SN--peak (due to SN~II) occurs depends on the exact
mass limit of WR star formation, assumed here to be $25$ \Msol.  E.g.\
lower values of $M_{\rm WR}$ should shift the SN~II peak to older ages.  A
similar temporal behaviour of \al\ is found in the models of \cite{Pl00}.

For \fe, the emissivity is composed of two broad bumps (at $\sim7$ and
$\sim17$ Myr) that are separated by a local minimum around $\sim13$ Myr.
After the second bump, the emissivity decreases smoothly, followed by the
exponential decline after $\sim 37$ Myr.  Overall, the \fe\ production
stays almost constant between $5-18$ Myr, followed by a slow decline, and
due to the uncertainties in the nucleosynthesis calculations, we should
only retain this behaviour as characteristic.  Given our detailed treatment
of yields from SN~II and SN~Ib, a larger
\fe\ yield is obtained when compared to the models of 
\cite{Pl00}. 
This also affects the predicted \fe/\al\ ratio.

\subsection{$\fe / \al$ emissivity ratio}

\begin{figure}
\epsfig{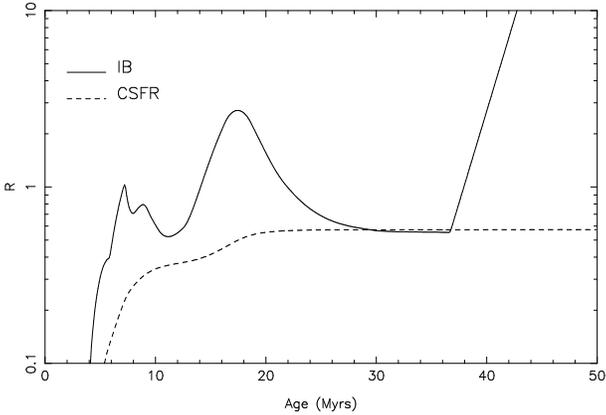}
\caption[]{\fe\ over \al\ emissivity ratio $R$ for the IB (solid) 
           and the CSFR model (dashed).  The steady state
           ratio amounts to 0.6.  }
\label{fig:FeAl}
\end{figure}

Part of the \al\ and \fe\ are co--produced in the same regions within type
II supernovae (Timmes et al.~1995\nocite{Timetal95}), hence the observation
of gamma--ray lines from both isotopes can be used as powerful diagnostics
tool of nucleosynthesis conditions in such events.  For this reason, we
investigate also the time--dependency of the ratio $R$ between \fe\ and
\al\ emissivities defined as $R(t) = \dot{Y}_{60}(t) / \dot{Y}_{26}(t)$.
The predicted dependency is shown in Fig.~\ref{fig:FeAl} for the case of an
instantaneous burst, and for a continuous star formation rate.

For the IB model, $R$ rises steadily as function of time.
This is due to the fact that \al\ ejection decreases with increasing 
age, while the \fe\ yields stay roughly constant during the active 
phase of the association.
After the last supernova exploded, at an age around $\sim37$ Myr,
the accumulated yields decay exponentially, leading to an exponential 
rise in $R$ with a time scale of
\begin{equation}
 \tau = \left( \frac{1}{\tau_{26}} - \frac{1}{\tau_{60}} 
 \right)^{-1} = 2.09 \, \,{\rm Myr}
\end{equation}
($\tau_{26}=1.04$ Myr and $\tau_{60}=2.07$ Myr are the mean lifetimes 
of \al\ and \fe, respectively).
Note that for ages younger than $\sim5$ Myr, copious \al\ production 
may appear in an association while no \fe\ has been synthesised yet.
This is due to the fact that no appreciable amounts of \fe\ are supposed 
to be ejected by stellar winds, and \fe\ appears only when the 
first supernovae begin to explode.
Again, this should be a generic feature of the time--evolution of a 
coevally formed population that should be independent of details in 
the nucleosynthesis calculations.

For a constant star formation rate, $R$ rises more smoothly and soon 
turns into its steady state value around $0.6$.
Our result is slightly in excess of the calculations of \cite{Timetal95} 
who inferred a value of $R=0.38 \pm 0.27$ from a chemical evolution 
calculation for the Galaxy, assuming only SN~II nucleosynthesis 
without mass--loss in the initial mass range $11-40$ ~{\Msol} 
and taking the metallicity gradient of the Galaxy into account. 
Adopting similar assumptions\footnote{Stellar models with no mass loss,
\al\ produced exclusively in SNII with a mass range $11-40$~{\Msol},
Salpeter IMF from 0.08 to 40 {\Msol}.}
we obtain a ratio of $R=0.43$, which is much closer to their findings.
The main difference between the \cite{Timetal95} and our work 
lies in the treatment of mass loss during the hydrostatic burning
phases and its effect on the presupernova structure, which leads
to a reduction of \al\ by about 30\%, while the \fe\ nucleosynthesis 
remains almost the same.

\subsection{Equivalent O7~V star yields}

\begin{figure*}[ht]
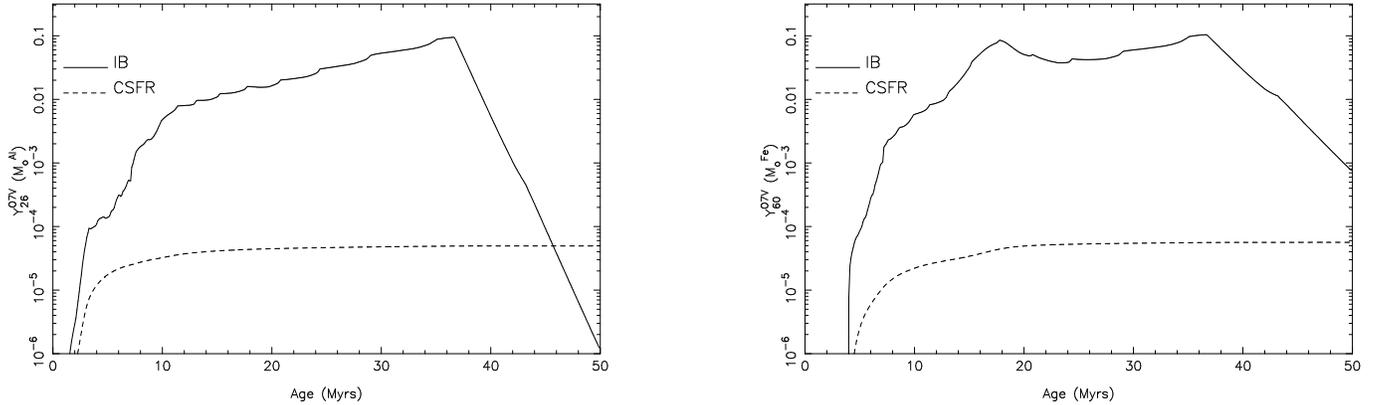

\epsfig{file=10196a.f8, angle=270,  width=8cm}
\hfill
\epsfig{file=10196b.f8, angle=270,  width=8cm}
\caption[]{Equivalent O7~V star \al\ yield (left) and \fe\ yield 
           (right) as predicted by the IB (solid) and the CSFR (dashed)
           model. 
}
\label{fig:YO7}
\end{figure*}

COMPTEL observations of the 1.809 \MeV\ gamma--ray line suggest that the 
\al\ flux
is proportional to the number of ionising photons 
(Kn\"odlseder et al.~1999\nocite{knoedl99a}).
In order to express the proportionality factor in convenient units, 
\cite{knoedl99} introduced the ``equivalent O7~V star \al\ yield'', \yal,
as the mass of \al\ that is produced by a star of spectral type O7~V, 
assuming that such a star has an ionising flux of $\log Q_0^{\rm O7~V}$ = 
49.05 ph s$^{-1}$.
This terminology follows closely the one employed for the analysis of 
starburst galaxies, where the strength of the ionising flux is often 
expressed in terms of equivalent stars of a given subtype 
(e.g.~Vacca 1994\nocite{vacca94}).
We extend here the definition also to \fe, where in analogy \yfe\ is 
the mass of \fe\ produced by an O7~V star.

The predicted equivalent O7~V star yields, calculated using
\begin{equation}
 Y^{\rm O7~V}(t) = \frac{10^{49.05}}{Q_{0}(t)} \times \dot{Y}(t)  
 \times \tau ,
\end{equation}
are shown in Fig.~\ref{fig:YO7} for the IB and the CSFR model.
Apparently, \yal\ is an extremely sensitive age indicator for a coevally 
formed population (this would be also true for \yfe, yet the \fe\ lines 
have not been detected so far).

Within 10 Myr, \yal\ varies by more than 3 orders of magnitude.  This
strong variation is mainly due to the rapid evolution of the ionising flux
which drops considerably as soon as the most massive stars in the
association vanished (see Section \ref{sec:Q}).  In comparison, the most
common age--indicator in \HII\ regions, the H$\beta$ equivalent width,
varies only within 3 orders of magnitude within 20 Myr (see Cervi\~no \&
Mas-Hesse 1994\nocite{CMH94} for more details).  A combination of radio
observations providing ionising fluxes and 1.809 \MeV\ gamma--ray
observations should allow to obtain age estimates.
In particular, although the strong time-variation is driven by 
the rapid drop in ionising flux, adding the gamma-ray observations provides 
a convenient normalisation, making the age estimate independent of 
distance, population richness, interstellar extinction, and IMF 
slope (see also Sect.~\ref{sec:IMFdep}).

The evolution of \yal\ for the IB model can be split into four phases:
\begin{enumerate}
\item {\em the stellar wind phase}, lasting from the star formation burst 
      up to $3$ Myr, and which is characterised by a steep rise of 
      \yal,
\item {\em the type Ib/c supernova phase}, from $3-7$ Myr, showing a 
      flattening in the slope, which comes from a slight decline in \al\ 
      production together with a rapid decline in the ionising flux,
\item {\em the type II supernova phase}, from $7-37$ Myr, 
      starting with a step around $7$ Myr due 
      to the most massive SN~II, followed by a tail of positive slope 
      since the ionising flux drops quicker than \al\ production, and
\item {\em the decay phase}, after $37$ Myr, which is dominated by 
      the exponential decay of \al.
\end{enumerate}
While this general picture should not depend on details of the 
nucleosynthesis and atmosphere models, the exact slopes and time 
intervals may well change for different input physics.
In particular, phase 3 could already start after $5$ Myr if the 
minimum mass required to form a Wolf--Rayet star would be as high as 
$40$ \Msol.
Figure \ref{fig:YO7} also illustrates that the equivalent \al\ 
star yield is an excellent discriminator between O star 
nucleosynthesis (i.e.~hydrostatic nucleosynthesis in WR stars and 
explosive nucleosynthesis in type Ib/c SNe) and SN~II
nucleosynthesis.
While phase 1 and 2 are characterised by low \yal\ values, ranging 
from zero to $\sim5 \times 10^{-4}$ \Msol, SN~II phase 
is characterised by high equivalent yields well above $\sim10^{-3}$ \Msol.
Due to the order of magnitude difference, it should be relatively easy 
to use \yal\ to discriminate between both contributions.

Although the evolution of \yfe\ for the IB model follows roughly that 
of \yal, we cannot easily identify distinct phases as for \al.
Due to the complex behaviour of the \fe\ yields, a lot of structure is found
in the evolution of the \fe\ ejection rates, but no simple characteristic 
trend.  Hence, to first order, the evolution of \yfe\ mainly 
reflects the fast decline of the ionising flux with increasing age.

\begin{table}[th]
  \caption{\label{tab:steadystate}
  Steady--state predictions of the equivalent O7~V star yields.}
  \begin{flushleft}
    \begin{tabular}{llcc}
    \hline
    \noalign{\smallskip}
    & Source & $Y^{\rm O7~V}$ (\Msol) & Contribution (\%) \\
    \noalign{\smallskip}
    \hline
    \noalign{\smallskip}
    \yal\ & MS--winds & $0.46 \times 10^{-5}$ &   9 \\
          & WR--winds & $1.60 \times 10^{-5}$ &  33 \\
          & SN~Ib/c  & $0.69 \times 10^{-5}$ &  14 \\
          & SN~II    & $2.19 \times 10^{-5}$ &  44 \\
    \noalign{\smallskip}
    \cline{2-4}
    \noalign{\smallskip}
          & total    & $4.94 \times 10^{-5}$ & 100 \\
          &          &                       &     \\
    \yfe\ & SN~Ib/c  & $2.18 \times 10^{-5}$ &  39 \\
          & SN~II    & $3.45 \times 10^{-5}$ &  61 \\
    \noalign{\smallskip}
    \cline{2-4}
    \noalign{\smallskip}
          & total    & $5.63 \times 10^{-5}$ & 100 \\
    \noalign{\smallskip}
    \hline
    \end{tabular}
  \end{flushleft}
\end{table}

The models calculated for a constant star formation rate allow us to 
predict steady state equivalent yields, which are reached after 
$\sim20$ Myr (cf.~Fig.~\ref{fig:YO7}).
The resulting steady state values, split into contributions from 
individual source types, are given in Table \ref{tab:steadystate}.
In particular, stellar wind contributions have been divided into
yields ejected prior to (MS--winds) and during (WR--winds) the Wolf--Rayet 
phase, respectively.
Apparently, $1/4$ of the hydrostatically produced \al\ ejected by stellar 
winds comes from before the WR phase, while the rest is ejected 
when the Hydrogen envelope gets entirely lost in a Wolf--Rayet phase.
As already pointed out by MAPP97\nocite{MAPP97} and \cite{knoedl99}, 
stellar wind ejection from massive stars provide an important 
($\sim$ 42 \%) contribution to the global \al\ production.
Type II supernovae contribute a similar amount, while the rest originates
from SN~Ib/c explosions.
The exact repartition on the different source classes depends, of 
course, on the nucleosynthesis models, but also on the assumed mass 
limit for WR star formation, the slope of the IMF, and finally the 
metallicity (Kn\"odlseder 1999\nocite{knoedl99}).

Our model predicts a steady--state equivalent O7~V star \al\ yield of 
$4.94 \times 10^{-5}$ \Msol , which is 
lower than the observed value integrated over the whole Galaxy
of $(1.0 \pm 0.3) \times 10^{-4}$ \Msol\
(Kn\"odlseder 1999\nocite{knoedl99}).
In view of the uncertainties involved in the nucleosynthesis 
calculations, the similarity between model and observation is however
encouraging.
In addition, our models were calculated for solar metallicity only, 
whereas the gamma--ray observations average over the entire Galaxy, 
which shows an average metallicity of roughly twice the solar value 
(Prantzos \& Diehl 1996\nocite{prantzos96}).
Higher metallicities potentially increase the \al\ production by 
Wolf--Rayet stars, due to an increase in mass--loss and the amount of seed 
nuclei available for \al\ synthesis (e.g.~MAPP97\nocite{MAPP97}).
Hence, including metallicity effects in our calculations is expected 
to raise the \yal\ estimate, bringing it even closer to the observed 
value.

\begin{table}[th]
  \caption{\label{tab:galactic}
  Galactic yield predictions assuming solar metallicity derived in 
  this work, and given by Kn\"odlseder (1999). The star formation 
  rate (SFR) is quoted for the mass interval $1-120$~\Msol in  
  Kn\"odlseder (1999) work.}
  \begin{flushleft}
    \begin{tabular}{lcc}
    \hline
    \noalign{\smallskip}
    & this work & Kn\"odlseder (1999) \\
    \noalign{\smallskip}
    \hline
    \noalign{\smallskip}
    $\dot{Y}_{26}$ (\Msol\ Myr$^{-1}$) & 1.47 & $1.5 \pm 0.3$ \\
    $\dot{Y}_{60}$ (\Msol\ Myr$^{-1}$) & 0.84 & - \\
    $M_{26}$ (\Msol)                   & 1.53 & $1.6 \pm 0.3$ \\
    $M_{60}$ (\Msol)                   & 1.74 & - \\
    SN rate (SN century$^{-1}$)        & 2.44  & - \\
    SFR (\Msol\ yr$^{-1}$)             & 0.96 & 1.2 \\
    \noalign{\smallskip}
    \hline
    \end{tabular}
  \end{flushleft}
\end{table}

Using the estimated galactic Lyman continuum luminosity of 
$Q = 3.5 \times 10^{53}$ photons s$^{-1}$
(Bennett et al.~1994\nocite{bennett94}), the number of equivalent 
O7~V stars can be estimated to $31\,194$, and we can predict galactic 
nucleosynthesis yields from our CSFR model.
A similar approach has been followed by \cite{knoedl99} using a 
time--independent steady--state model for the Galaxy. 
In Table \ref{tab:galactic} we compare his 
findings for solar metallicity and Salpeter IMF with mass limits 1 -- 120 
{\Msol}, to our model 
(Salpeter IMF with mass limits 2 -- 120 {\Msol}) 
and fitting the ionising flux to the observed value.
Overall, the agreement between the models is quite satisfactory.
Our models predict a total Galactic \fe\ mass of 1.7 \Msol , which
due to cancellation of various differences, turns out to be very 
similar to the value of \cite{TW97}.

\subsection{Dependence on IMF slope}
\label{sec:IMFdep}
No general consensus exists about the slope of the IMF in young 
massive star associations and related objects (see reviews in
Gilmore \& Howell 1998). 
For example from an analysis of young open clusters and OB association in the 
Milky Way, \cite{Mass95} derive an average slope of 
$\Gamma=-1.1 \pm 0.1$ for stars with masses $> 7 \Msol$.
For O stars within 2.5 kpc from the Sun \cite{Garmany82}  find $\Gamma=-1.6$. 
Based on NIR photometry of the massive Cyg OB2 association, 
\cite{knoedl00} found a comparable slope of $\Gamma=-1.6 \pm 0.1$.
Finally, in his most recent revision \cite{krou00}, obtains
$\Gamma=-1.3 \pm 0.7$ for stars with masses $> 1 \Msol$, taking
the scatter introduced by Poisson noise and the dynamical evolution of 
star clusters into account.

Throughout this work a Salpeter IMF slope ($\Gamma = -1.35$) has been
used for our ``standard'' models. The dependence of our results on
$\Gamma$ are illustrated subsequently.

\begin{figure}[ht]
\epsfig{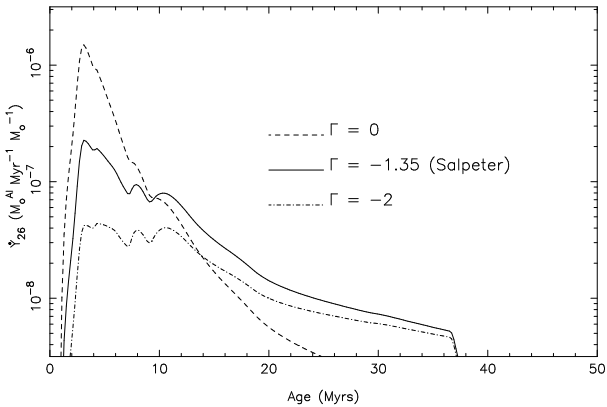}
\caption[]{\al\ emissivity for three different IMF slopes.
}
\label{fig:alejIMF}
\end{figure}

Fig.~\ref{fig:alejIMF} shows the time-dependent \al\ emissivity for 
assuming an IMF slope of $\Gamma = 0.0$, $-1.35$, and $-2.0$. 
All three curves have been normalised to the mass transformed into stars in
the mass range 2 -- 120 M$_\odot$
Obviously, the structure in the time-evolution remains similar, but 
the importance of stellar wind ejecta with respect to supernova 
ejecta depends strongly on $\Gamma$.
For $\Gamma=0.0$ the stellar-wind \al\ emissivity peak (at $\sim 3$ 
Myr) is almost one magnitude larger than for the Salpeter law, leading 
to a burst-like lightcurve that is dominated by stellar wind products.
In contrast, for $\Gamma=-2.0$ the stellar-wind emissivity is of the 
same level as the type II supernova emissivity, leading to an almost 
10 Myrs lasting plateau in the lightcurve.

\begin{figure*}[ht]
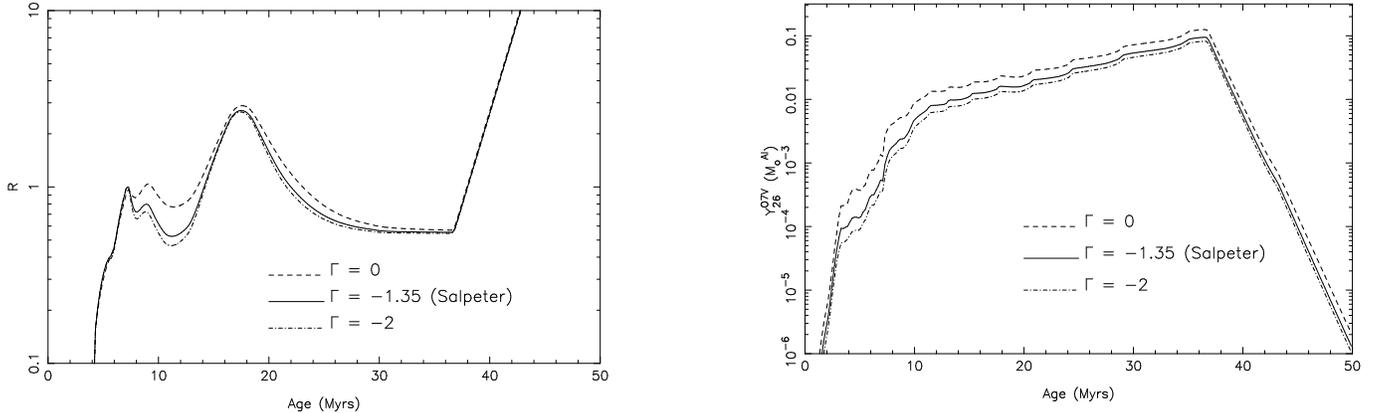

\epsfig{file=10196a.f10, angle=270,  width=8cm}
\hfill
\epsfig{file=10196b.f10, angle=270,  width=8cm}
\caption[]{$\fe / \al$ emissivity ratio $R$ (left) and equivalent O7~V 
           star \al\ yield (right) for three different IMF slopes.
}
\label{fig:IMF}
\end{figure*}

Interestingly the $\fe / \al$ emissivity ratio $R$ and the equivalent
O7~V star \al\ yield \yal\ depend very little on the IMF slope, as
shown in Fig.~\ref{fig:IMF}. 
This is due to the fact that both the nucleosynthetic yield and the
ionising flux show a similar dependence with initial mass.
This finding indicates that the equivalent O7~V star \al\ yield should
be fairly reliable age indicator for young massive star associations.
From Fig.~\ref{fig:IMF} we estimate a typical age uncertainty of 
$\pm2$ Myr due to IMF variations, which is of the same order as 
typical uncertainties obtained for massive star associations by isochrone 
fitting.
Also, the IMF variations are smaller than the dispersion introduced by 
statistical fluctuations in a finite sample, as we will demonstrate 
for realistic populations in section 5 (cf.~Fig.~\ref{fig:pdfs} right 
panel).

\section{Model uncertainties}
\label{sec:uncertain}

Before we proceed to the description of our technique which will
be used in future applications to compare our model predictions
with observations we shall briefly discuss potential uncertainties
from stellar models which may affect our results.
Uncertainties arising from the choice of nucleosynthesis models 
have been discussed earlier (Sect.\ \ref{sec:yields}) and shall not
be repeated here.


The main uncertainties related to the stellar models included in our
synthesis are likely the neglect of stellar rotation and the possible
importance of massive close binary stars as sources of \al\ production.

\subsection{Rotation}
Rotation induces numerous dynamical instabilities in stellar interiors.
The related mixing of the chemical species can deeply modify the chemical
structure of a star and its evolution (see the review by Maeder \& Meynet
2000\nocite{Ma20}), and may in particular have important consequences on
the $^{26}$Al production by WR stars (cf.\ Arnould et
al.~1999\nocite{Arnetal99}). At present, very few rotating WR models
address this question in detail (see Langer et al.~1995\nocite{langer95}
for some preliminary results), so that it is certainly premature to
quantitatively assess the possible role rotation plays in that
respect. However, it seems safe to say that rotation increases the quantity
of $^{26}$Al ejected by WR stellar winds (see Arnould et al.\
1999\nocite{Arnetal99}).  This statement is corroborated by the numerical
simulations of \cite{langer95} which show that in case of fast rotation,
\al\ production may be enhanced by factors of 2 to 3 with respect to
non--rotating stars.  An increase of \al\ by a factor of $\sim$ 1.5 over the
non--rotating case was found recently by Ringger (2000)\nocite{Rin20} for a
60 M$_\odot$ model with an average rotational velocity during the Main
Sequence of 170 km/s.

Rotation also lowers the
minimum initial mass of single stars which can go through a WR stage.
It is thus likely to increase the net
amount of \al\ ejected by WR stars in the ISM. 
However, since in the present work we use stellar tracks with enhanced mass 
loss rates which, to a certain extent, mimic some effects of rotation,
it may be expected that the inclusion of rotation will not drastically
alter the present results.

\subsection{Binaries}

The case of binaries is more complicated. First of all the stellar physics 
is even more complex than in single stars, which adds numerous potential
uncertainties. Second the frequency of interacting massive close binaries
and their exact scenario is poorly known.
The importance of binaries in evolutionary synthesis models is therefore
still difficult to assess (e.g.\ Mas-Hesse \& Cervi\~no 1999\nocite{MHC99}, Vanveberen 1999\nocite{vanveberen99}).

For the case of binary systems two effects must be taken into
account: {\em 1)} tidal interactions in close binary systems,
and {\em 2)} mass transfer by Roche Lobe Overflow and the posterior 
evolution of the primary and secondary stars.
Before any mass transfer will occur tidal effects are expected to deform the 
star and therefore induce instabilities reminiscent of those induced by rotation. 
To our knowledge, the latter effect has never been studied, even if it might have 
important consequences, as for instance by homogenizing the stars and thus
inhibiting any mass transfer. 
If mass transfer occurs, the removal of part of the envelope of the donor star 
may favour the  appearance of \al\ on the surface in a similar fashion as mass 
loss through stellar winds.
More important changes may occur for the gainer in systems with
primaries $M_1 \la 40$ \Msol, as e.g.\ shown
by the preliminary studies of \cite{Br95} and \cite{langer98}.
The latter suggest for example a scenario in which mass transfer onto the secondary 
leads to a rejuvenation which alters its subsequent evolution.
For this case they predict an increase by $2-3$ orders of magnitudes of the 
hydrostatically produced \al\ yield due to a reduction of the delay between 
production and ejection of \al, which could possibly enhance the total \al\ 
production from type II supernovae by a factor of about 2 
(Langer, priv.~communication).

An attempt to include the contribution of binaries to the production
of \al\ by a stellar population was presented by \cite{Pl00}.
Given the limited knowledge on the evolution and nucleosynthesis from 
binary systems and the numerous potential uncertainties affecting these 
predictions, their contribution has deliberately been neglected in the 
present models. As for rotation, if all possible effects of binarity 
produce larger  amounts of \al, our model predictions represent a lower 
limit for the \al\ production.

\section{Realistic populations}
\label{sec:realistic}

The model predictions presented in Section \ref{sec:predictions} have 
been calculated for an infinitely rich association of stars in order 
to avoid effects due to small number statistics, in particular at the 
high--mass end of the IMF.
In reality, however, galactic associations or open clusters have total 
stellar masses between a few $100-1000$ \Msol\ (Bruch \& Sanders 
1983\nocite{bruch83}), limiting the number of O stars, i.e.~stars 
with initial masses above $\sim20$ \Msol, to only a few objects.
However, since these massive stars provide a considerable 
fraction of the nucleosynthesis yields and ionising power of the 
association, the early evolution of the system will depend rather 
sensitively on the actual distribution of stellar masses.
In the following we will describe a  Bayesian method that allows to quantify 
the uncertainties introduced by a finite sample in our predictions.

\subsection{Probability density functions}

\begin{figure*}[ht]
\epsfig{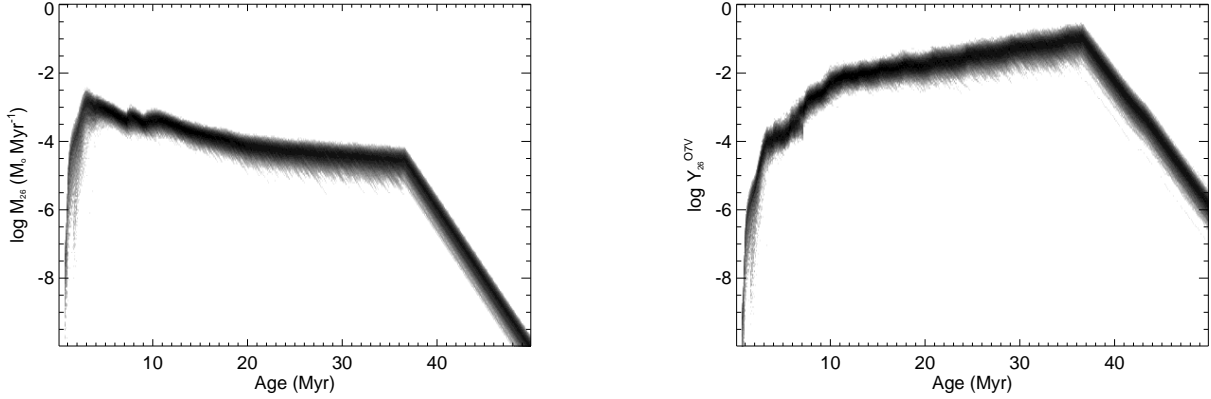}
\caption[]{Time--dependent probability density functions for the 
           \al\ emissivity (left) and \yal\ (right).
           A logarithmic greyscale was chosen to display also the 
           wings of the PDFs.}
\label{fig:pdfs}
\end{figure*}

Suppose we want to predict the gamma--ray luminosity and ionising flux 
from an 
association of age $5\pm1$ Myr, which 
today contains 100 stars within an initial mass range of $8-25$ \Msol.
At 5 Myr, stars above $\sim 40$ \Msol\ already exploded as supernovae, 
but the population within $8-25$ \Msol\ should still be relatively 
unaffected by evolution.
A possible initial population may then be estimated by randomly 
selecting stellar masses from a power--law IMF of slope $\Gamma$ until 
the number of stars with masses between $8-25$ \Msol\ amounts to 100
(cf.~section \ref{sec:method}). 
This population is then statistically identical to the initial 
population of the observed association, although it may differ in the 
exact distribution of stellar masses.
Repeating the sampling provides 
several possible initial realisations, 
and the statistical variations in the number and distribution of 
stars among these samples reflects then the ignorance about the 
precise initial conditions in the association.

The calculation of evolutionary synthesis models for each of the samples 
provides then time--dependent model predictions, and the variations 
among the predictions reflect the uncertainty about the precise 
distribution of the initial stellar population.
If the number of random samples and corresponding evolution synthesis 
models is sufficiently high
the probability 
$p(x|t)$ of observing at an age $t$ the value $x$ for quantity $X$ can be 
reasonably well approximated by 
\begin{equation}
 p(x|t) \Delta x \approx \frac{N(x \le X < x + \Delta x)}{N_{\rm samples}} ,
\end{equation}
where $N(x \le X < x + \Delta x)$ is the number of samples for which $X$ 
was comprised between $x$ and $x + \Delta x$, and $N_{\rm samples}$ is 
the total number of samples.
$p(x|t)$ is called the probability density function (PDF) of $X$ at 
age $t$.

For illustration, time--dependent probability density functions for the 
\al\ emissivity and \yal\ are shown for our example association 
in Fig.~\ref{fig:pdfs}.  The PDFs have been computed from 1000 Monte Carlo
samples assuming an IMF slope of $\Gamma=-1.35$.  During the early
evolution ($< 3-5$ Myrs), both quantities are subject to considerable
uncertainties which can be understood as the combined effect of comparable
lifetime, \al\ yield mass dependence, and small number statistics at the
high--mass end.  As mentioned earlier, stars with initial mass between
$60-120$ \Msol\ evolve on comparable timescales, providing contemporaneous
\al\ ejection in this mass range.
\al\ yields, however, vary by almost one magnitude within this mass 
range, making the resulting \al\ ejection rates crucially dependent on the
particular spectrum of initial masses.  Consequently, the large statistical
uncertainties in the mass spectrum at the high--mass end translates into a
large uncertainty in the \al\ emissivity and \yal\ during the early
evolution of the population.

During the subsequent evolution, the relative uncertainty in the derived
quantities is roughly constant, with a minimum around $5-10$ Myr followed
by a slight rise of the uncertainty with time.  To understand this
behaviour, note that the relative uncertainty in the number $n$ of stars
that contribute to \al\ production within a time step is given by
$n^{-1/2}$, hence the uncertainty increases with decreasing $n$.  Indeed,
for ages $>7$ Myr the \al\ production is only due to SN explosions, and the
slight decrease in the SN rate with time (Fig. \ref{fig:SNr}) is at the
origin of the uncertainty increase.  Note that this feature depends on the
slope $\Gamma$ that is chosen for the stellar population: with a steeper
IMF the decline in the supernova rate would have been reduced (or may even
turn into an increase), and consequently the uncertainty increase would
become negligible (or might even turn into a decrease of the
uncertainties).

\subsection{Inclusion of uncertainties}

\begin{figure*}[ht]
\epsfig{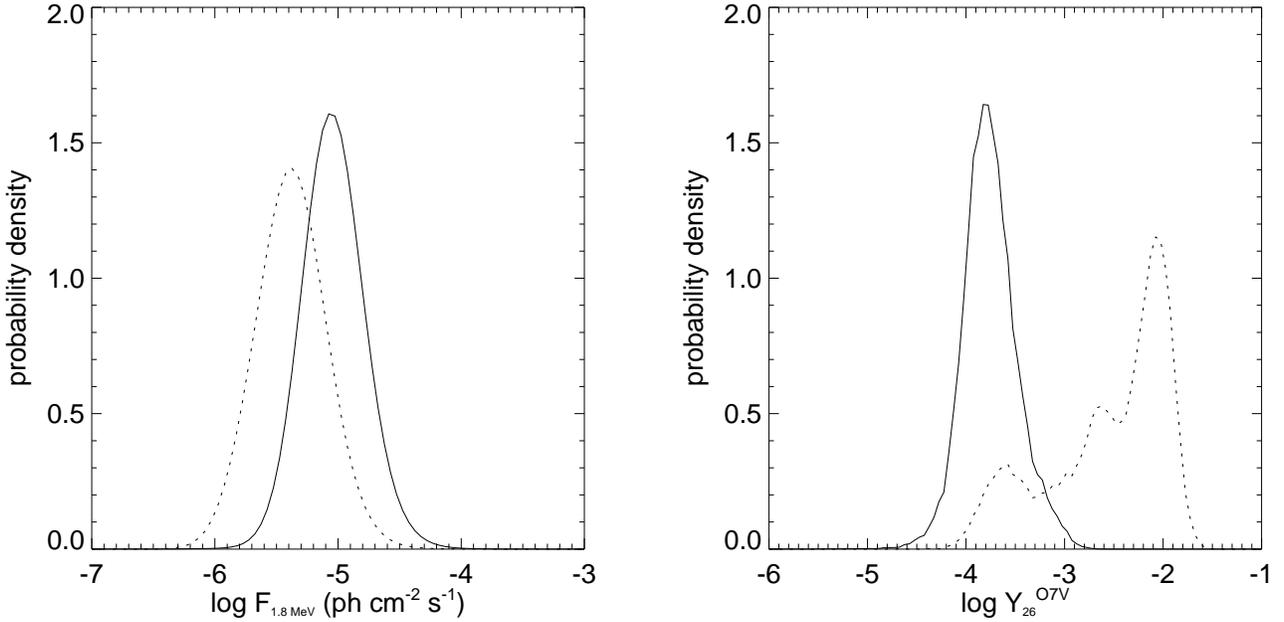}
\caption[]{Posterior PDFs for the 1.809 MeV gamma--ray flux from the 
           decay of radioactive \al (left) and the equivalent O7~V 
           star \al\ yield \yal\ (right).
           The solid lines show the results for an age of $5\pm1$ 
           Myr, the dashed line has been calculated for an age 
           uncertainty of $5-15$ Myr.
           A distance of $1.0\pm0.2$ kpc has been assumed for the association.}
\label{fig:postpdfs}
\end{figure*}

In the above example we assumed that the number of stars within a 
given mass interval has been determined precisely, and that the slope 
$\Gamma$ of the association is known.
Both assumptions may not be valid in a realistic case.
The determination of the association richness is certainly subject to 
some uncertainty, due to membership ambiguities, stellar confusion, 
or invisible members hidden by interstellar obscuration.

Uncertainties in both quantities can be easily included in the 
analysis by computing the conditional probability density function
$p(x|t,n,\Gamma)$ for a sufficiently fine grid of stellar richness 
$n$ and IMF slope $\Gamma$.
The time--dependent association PDF is then calculated using
\begin{equation}
 p(x|t) = \int p(n) p(\Gamma) p(x|t,n,\Gamma) dn d\Gamma ,
\end{equation}
where $p(n)$ and $p(\Gamma)$ are prior probability density 
functions quantifying the uncertainties in $n$ and 
$\Gamma$\footnote{This technique of removing irrelevant parameters 
from the problem is called {\em marginalisation}.}.
Typically, the prior PDF could be a Gaussian if the uncertainty is 
symmetric around a mean value, or a bounded constant if the 
uncertainty is specified as lower and upper limits.

\subsection{Inclusion of prior knowledge}

Finally, to predict the characteristics of an association, the 
information about its actual age and its distance should be included 
in the analysis.
In the above example, the age was estimated to $5\pm1$ Myr, which 
again can be expressed by a prior PDF $p(t)$.
Marginalisation leads then to an age independent estimate
\begin{equation}
 p(x) = \int p(t) p(x|t) dt .
 \label{eq:age}
\end{equation}
Equation \ref{eq:age} can be seen as a smoothing operation applied to the 
time--dependent PDF over a limited age--window, specified by the width of 
the prior PDF.
Since our models are calculated for instantaneous star formation, 
Eq. \ref{eq:age} can also be interpreted as extending the star 
formation to a period which is defined by the width of the prior PDF.
In this case, $p(t)$ is the star formation law, and the age 
uncertainty is interpreted as an uncertainty in the star formation 
history.

Similarly, if the distance $s$ of the association is known (to some 
uncertainty -- of course), yields and ionising flux can be converted to 
gamma--ray and radio fluxes, using
\begin{equation}
 p(\Phi) = \int p(s) p(\Phi|s) ds
\end{equation}
where $p(s)$ is the prior PDF, and $p(\Phi|s) \propto p(x) s^{-2}$ is the 
distance dependent flux PDF.

Thus, the final outcome of the evolutionary synthesis model is not a 
single value, but a probability density function, also called the 
{\em posterior probability density function}, that reflects all 
uncertainties related to the investigated association.
Two examples of such posterior PDFs are shown in Fig.~\ref{fig:postpdfs} 
for the 1.809 MeV gamma--ray flux and the equivalent O7~V star \al\ yield
\yal.
The first example (solid lines) has been obtained by assuming an age
uncertainty of $5\pm1$ Myr, expressed by a Gaussian prior PDF with mean of
5 Myr and standard deviation of 1 Myr.  The second example (dashed lines)
has been derived for an age uncertainty of $5-15$ Myr, described by a
constant bounded prior PDF which is non--zero for the interval $5-15$ Myr.
To obtain the 1.809 MeV flux from the \al\ emissivity a distance of
$1.0\pm0.2$ kpc has been assumed, implemented as a prior PDF of Gaussian
shape with mean of $1$ kpc and standard deviation of $0.2$ kpc.  Note that
\yal\ is not subject to any distance uncertainty since it is defined as a
flux (or yield) ratio for which the actual distance cancels out.

In many cases, the resulting posterior PDF has approximately a 
Gaussian shape, but the example of \yal\ using an age uncertainty of 
$5-15$ Myr illustrates that the distribution may be much more complex.
The posterior PDFs may then be used to address various questions about 
the investigated association.
For example, the probability $P$ that the 1.809 MeV flux is comprised in the 
interval $[\Phi_{\rm min},\Phi_{\rm max}]$ is simply derived by 
integrating
\begin{equation}
 P = \int_{\Phi_{\rm min}}^{\Phi_{\rm max}} p(\Phi) d\Phi .
\end{equation}
Or, inverting the problem, one may derive the flux interval
$[\Phi_{\rm min},\Phi_{\rm max}]$ that contains $68\%$ of the probability 
distribution by solving
\begin{equation}
 \int_{-\infty}^{\Phi_{\rm min}} p(\Phi) d\Phi = 0.16
\end{equation}
and
\begin{equation}
 \int_{-\infty}^{\Phi_{\rm max}} p(\Phi) d\Phi = 0.84 .
\end{equation}
(this definition is similar to the $1 \sigma$ errors often quoted in 
classical statistics).

The posterior PDFs may also be used to make predictions about the 
detectibility of an association by a gamma--ray instrument or 
radio telescope.
If the instrument sensitivity is given as smallest detectible flux 
limit $\Phi_{3 \sigma}$, the probability of detecting the association
with the instrument is given by
\begin{equation}
 P_{\rm detect} = \int_{\Phi_{3 \sigma}}^{\infty} p(\Phi) d\Phi .
\end{equation}

\section{Conclusions}

We have constructed a new diagnostic tool for the study of the radioactive
isotopes of \al\ and \fe\ produced in massive star forming regions. 
The main aim of this work was to provide a quantitative model for
the analysis of multi--wavelength observations of OB associations,
open clusters and alike objects covering the range from 
gamma--rays (e.g.\ the 1.809 MeV line of \al, 1.173 and 1.333 Mev lines
of \fe) to radio, and allowing in a fully quantitative manner
to account for statistical richness effects of massive star populations
and other observational uncertainties.

To achieve this goal we have used the evolutionary synthesis models
of \cite{CMH94}, 
which have been updated to include
recent Geneva stellar evolution tracks, new stellar atmospheres
for OB and WR stars, and nucleosynthetic yields from massive stars 
during hydrostatic burning phases and explosive SNII and SNIb events
(see Sect.\ 2). In particular proper care was taken to 
combine the stellar models including mass loss with appropriate 
presupernova and SN models. 

The temporal evolution of the ejected quantity of \al\ and \fe\ produced 
by a coeval population, other observables like the total ionising flux
and the supernova rate, and derived properties is presented (Sect.\ 3).
This yields the following main results:
\begin{itemize}
\item The equivalent O7 V star \al\ yield (\yal), defined as the stellar yield 
      per ionising flux of an O7 V star (Kn\"odlseder 1999\nocite{knoedl99}), shows a particularly 
      strong time dependence where four main phases can be distinguished: 
      stellar wind dominated phase ($\la$ 3 Myr), 
      SN Ib dominated phase ($\sim$ 3--7 Myr), 
      SN II dominated phase ($\sim$ 7--37 Myr), 
      and exponential decay phase ($\ga$ 37 Myr). 
      The exact age range of each phase is 
      dependent on the evolutionary tracks and the lower mass limit of WR stars.
\item The use of \yal\ is a powerful tool to constrain the evolutionary 
      status of star forming regions. 
      This parameter is obtained from a combined observation of the 
      $\gamma$--ray line and the ionising flux (for example in form of 
      thermal free--free radio emission) of an association.
\item \fe\ production starts with a delay of $\sim$ 2 Myr with 
      respect to \al\ production.  The ratio of the \fe/\al\ emissivities
      is also an age indicator that constrains the contribution of
      explosive nucleosynthesis to the total \al\ production.
\end{itemize}

Calculations for a steady state population (constant star formation; Sect.\ 3.5) 
at solar metallicity predict the following relative contributions to the \al\ 
production: 
$\sim 9 \%$ from stars before the WR phase, 
$\sim 33 \%$ from WR stars, 
$\sim 14 \%$ from SN~Ib, 
and $\sim 44 \%$ from SN~II.
The large contribution from stellar wind ejection ($\sim$ 42 \%) confirms 
earlier studies of MAPP97\nocite{MAPP97} and \cite{knoedl99} using 
similar yields, who predict contributions of 20--70 \% and $\sim$ 40 \%
respectively.
For \fe\ we estimate that $\sim 39 \%$ are produced by SN~Ib while 
$\sim 61 \%$ come from SN~II.
Normalising on the total ionising flux of the Galaxy, we predict total 
production rates of $1.5 \Msol\ $Myr$^{-1}$ and $0.8 \Msol\ 
$Myr$^{-1}$ for \al\ and \fe, respectively.
This corresponds to 1.5 \Msol\ of \al\ and 1.7 \Msol\ of \fe\ in
 the present interstellar medium.

As for other chemical evolution models, our calculations depend directly
on the adopted nucleosynthetic yields, which are affected by considerable
uncertainties (see e.g.\ Prantzos 1999\nocite{Pran99}). The main uncertainties regarding
\al\ and \fe\ have been discussed in Sects.\ 2 and 4.
In fact important new insight especially on the physics of supernovae is 
expected from the study of radioactive isotopes such as \fe\ and $^{44}$Ti, 
which are synthesised in deep layers close to the so--called mass cut separating
the outer regions from the remnant.
Such studies should also benefit from the present models.

Last, but not least, we have presented a Bayesian approach to quantify the
predicted observables and their uncertainty related to richness effects
of the IMF in terms of probability density functions (Sect.\ \ref{sec:realistic}).
Subsequently these functions can be used in combination with prior
knowledge on observed objects (e.g.\ age, distance, and their uncertainties) 
to calculate detection probabilities and alike quantities.
We have already successfully applied our models to existing multi wavelength 
observations of the Cygnus and Vela regions. The results will be published
in companion papers (Kn\"odlseder et al., in preparation; Lavraud et 
al., in preparation).
Our tools will be ideal to fully exploit the gamma--ray line observations 
expected from the upcoming {\em INTEGRAL} satellite.

\begin{acknowledgements}
Financial support from the ``GdR Galaxies'' and the ``Programme National de 
Physique Stellaire'' is acknowledged.
We also acknowledge the comments form an anonymous referee.
MC is supported by an ESA fellowship.

\end{acknowledgements}


\end{document}